# TERNARY INSTANTANEOUS NOISE-BASED LOGIC

LASZLO B. KISH [1,2]

[1] *Department of Electrical and Computer Engineering, Texas A&M University, College Station, TX 77843-3128, USA*

[2] *Óbuda University, Budapest, Bécsi út 96/B, Budapest, H-1034, Hungary*

One of the possible representations of three-valued instantaneous noise-based logic is proposed. The third value is an uncertain bit value, which can be useful in artificial intelligence applications. There is a forth value, too, that can represent a non-existing bit (vacuum-state) that is the same (1 numeric value) for all bits, however that is a squeezed state common for all bits. Some logic gates are explored. A ternary Universe has a significant advantage compared to the standard binary one: its amplitude is never zero during any clock period. All the known binary logic gates work for the binary bit values in the same way as earlier therefore the former binary algorithms can be run in the ternary system with no change and without the problems posed by zero values of the Universe.

## 1. Introduction

*1.1. On noise-based logic*

Noise-based logic (NBL) [1,2], is a deterministic scheme where the logic information is represented by truly random, stationary, orthogonal stochastic processes (noises with zero mean). In binary NBL, a pair of noises represents the High (H, 1) and Low (L, 0) bit values of the *noise-bit*. For $M$ noise-bits, $2M$ orthogonal noises are required. The $2M$ noise-bit values are provided by the Reference Noise System (RNS) and they are distributed all over the NBL processor (see Figure 1). With the received input logic information (noises) and the reference noises, the NBL gates create their output signals, which are also noises. It can be: any of the reference signals; their arbitrary superposition; their products; or the arbitrary superposition of their products. The products of noises, which are strings representing binary numbers, makes an exponentially large, $2^M$ dimensional Hilbert space (hyperspace) [1,2].

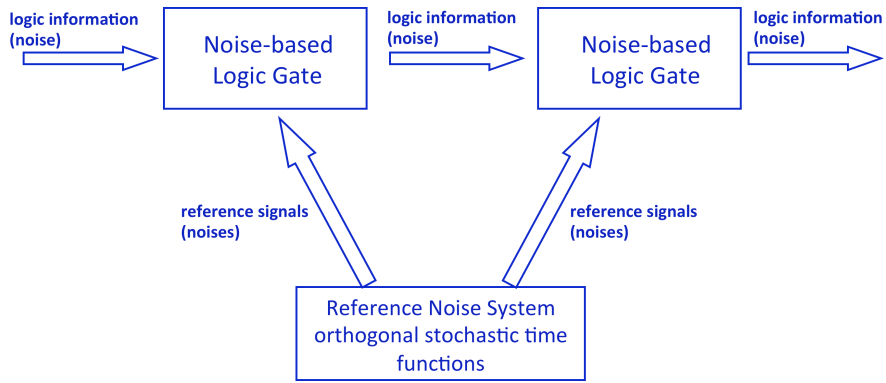

**Figure 1.** Generic noise-based logic hardware scheme [13]. Logic operations can be executed by the gates or by operations on the reference signals. The Reference Noise System is based on a truly random number generator.

*1.2 On binary instantaneous noise-based logic*

Instantaneous noise-based logic (INBL, [3-16]) does not require time averaging to create the output of the logic gate. It has many sub-types depending on the kind of the noise utilized. In the present paper, the reference noises have a periodic clocking and they are independent two-state noises, called random telegraph waves (RTW), with $R(t) = \pm 1$ amplitudes (each one with probability 0.5) for both the H (1) and the L (0) bit values, respectively, of a given noise bit [3,4]. The value is determined at the beginning of the clock period by a truly random coin. Such an RTW can be viewed as the binary fingerprint of the corresponding logic value of the given bit. As an example, see Figure 2.

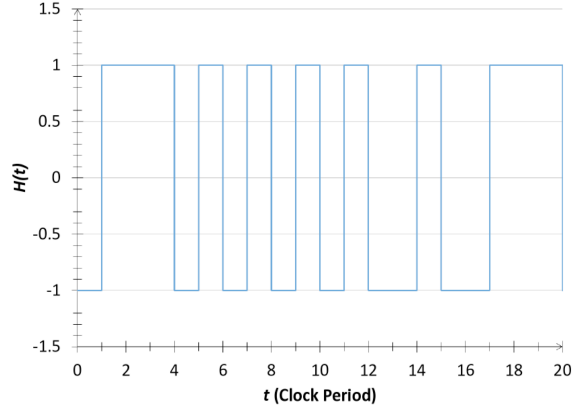

**Figure 2.** Example [11] of random telegraph wave (RTW) carrying a bit value of a chosen noise-bit in the asymmetric INBL scheme. The clock is periodic. At the beginning of each clock period, an unbiased, truly random coin generates the choice between the two possible amplitude levels. An *M*-noise-bit system requires 2*M* such RTWs.

The RNS of an *M*-noise-bits system with the 2*M* truly random noise generators is given as:

$$R_{10}(t), R_{11}(t), R_{20}(t), R_{21}(t), ..., R_{M0}(t), R_{M1}(t) \quad , \tag{1}$$

where the first index is the bit significance index and the second one represents the 0 (L) or 1 (H) values of that noise-bit. The reference noises (in the infinite-time limit) can also be viewed as *orthogonal vectors*:

$$\left\langle R_{ij}(t) R_{pq}(t) \right\rangle = 0 \quad \text{for} \quad i \neq p \ \text{ and/or } \ j \neq q \ , \tag{2}$$

where $\langle \ \rangle$ means infinite time average.

Thus, the superpositions of these signals represent a 2$M$ dimensional geometrical *space*.

However, product operations between the reference noises and their superpositions can lead out from this original space and form a *hyperspace* [1] which is also a Hilbert space, with exponentially high ($N=2^M$) dimensions representing an exponentially large number ($N$) of classical bits [2]. Figure 3 shows a circuit-block illustration of an INBL processor, which can be realized by a PC that can handle O($M$) long words (that is, having O($M$) bit resolution), and a truly random noise generator.

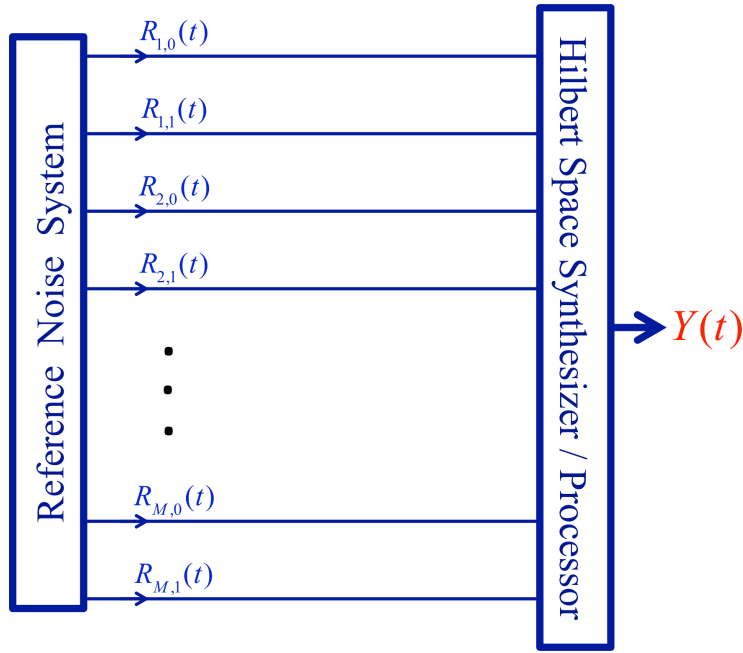

**Figure 3.** Circuit-block illustration of the logic structure of the generic superposition synthesizer for instantaneous NBL (INBL). The RNS contains 2$M$ independent RTW generators representing the $M$ noise-bits. The Hilbert Space Synthesizer is an algorithm that is typically containing multiplications, additions and subtractions. The output signal $Y(t)$ can represent a logic state with an exponentially large number of, O($N$), classical bit values. For examples, see the text.

Some examples:

In the INBL systems, binary numbers are $M$-bit long product-strings formed by the noises of the corresponding reference signals. For example, in a 3 noise-bit system, $M = 3$, the signal $W_6(t)$ of the number 6 and its binary version, 110, is carried by the noise product

$$W_6(t) = R_{10}(t)R_{21}(t)R_{31}(t) , \qquad (3)$$

where bit-1 is the bit of the lowest significance, and bit-3 is of the highest significance.

In a 4 noise-bit system, the signal $Y_{7,4,1}(t)$ of the superposition of the numbers 7 (111); 4 (100); and 1 (001) is:

$$\begin{aligned} Y_{7,4,1}(t) &= W_7(t) + W_4(t) + W_1(t) = \\ &= R_{11}(t)R_{21}(t)R_{31}(t) + R_{10}(t)R_{20}(t)R_{31}(t) + R_{11}(t)R_{20}(t)R_{30}(t) \end{aligned} . \qquad (4)$$

The product strings belonging to different numbers are orthogonal due to Equation (2), so this system is a Hilbert space with $2^M$ dimensions.

Some of the exponentially large, O($N=2^M$) superpositions of product strings can be set up by polynomial-in-$M$ complexity in INBL. For example, the Universe $U(t)$, which is the superposition of *all* the binary numbers in the $M$ noise-bit system, can be created by putting into superposition the L (0) and the H (1) bit values of each noise-bit and then multiplying these sums (Achilles heel operation [2,5,11,13]):

$$U(t) = \left[R_{10}(t) + R_{11}(t)\right]\left[R_{20}(t) + R_{21}(t)\right] ... \left[R_{M0}(t) + R_{M1}(t)\right] \qquad (5)$$

The time function *U*(*t*) of the Universe is set up by *M* additions and *M*-1 multiplications. When Equation (5) is expanded, it forms an exponentially large superposition, representing the superposition of all the binary numbers which this *M* noise-bit system can form. For example, in the 3 noise-bit case (*N*=8), it is:

$$\begin{aligned} U_{M=3}(t) = &R_{10}(t)R_{20}(t)R_{30}(t) + R_{11}(t)R_{20}(t)R_{30}(t) + R_{10}(t)R_{21}(t)R_{30}(t) + \\ &+ R_{11}(t)R_{21}(t)R_{30}(t) + R_{10}(t)R_{20}(t)R_{31}(t) + \\ &+ R_{11}(t)R_{20}(t)R_{31}(t) + R_{10}(t)R_{21}(t)R_{31}(t) + R_{11}(t)R_{21}(t)R_{31}(t) \ . \end{aligned} \quad (6)$$

*1.3 On ternary (trinary) logic*

Ternary logic [17,18] is a form of logic that allows for three bit values instead of the two classical values (false and true; or Low and High) used in binary logic. For example the three values used in ternary logic can be represented as Low (L, 0), High (H, 1), or Uncertain (X, 2).

The feature that ternary logic can represent uncertainty can be useful in applications where it is important to represent the possibility of error or uncertainty, such as in artificial intelligence.

In ternary logic, logical operations such as AND, OR, and NOT are defined to work with the three values. For example, the AND operation in ternary logic returns H only if both inputs are H, and L if either input is L. However, if both inputs are indeterminate (X), the result is also X.

Finally, ternary logic can be used to represent more complex relationships than binary logic. This is because there are more possible combinations of values in ternary logic than in binary logic. For example, in binary logic, the only possible relationship between two values is that they are either equal or unequal. In ternary logic, there are three possible relationships between two values: they can be equal, unequal, or unknown. This additional complexity can be useful in applications where it is important to represent relationships that are not simply equal or unequal, such as in natural language processing.

In conclusion, ternary logic is a useful tool that can be used in a variety of applications. It offers several advantages over binary logic, including the ability to represent values more precisely, uncertainty, and more complex relationships. However its classical hardware is more complex than that of Boolean logic. In the rest of the paper we introduce an INBL version of ternary logic that does not require significantly more complex hardware than that of binary INBL.

## 2. A ternary Instantaneous Noise-based Logic

Below, we show one of the possible realizations of ternary INBL (TINBL). It is a simple generalization of the binary INBL (BINBL).

*2.1 The basis of the TINBL*

The possible values of the *i*-th bit are:

**(a)** High: $H_i$ is the same as in BINBL, it is represented by an RTW, denoted as $R_{i1}(t)$

**(b)** Low: $L_i$ is the same as in BINBL, it is represented by another RTW, denoted as $R_{i0}(t)$

**(c)** Uncertain: $X_i$ is represented by the product of the above RTWs:

$$R_{ix}(t) = R_{i1}(t) R_{i0}(t) \tag{7}$$

Note, because $R_{ix}(t)$ is an unknown bit value, it has 1 bit information entropy.

**(d)** Non-existent bit: The amplitude value 1 instead of $R_{ix}(t)$ or $R_{i0}(t)R_{i1}(t)$ at bit *i* is a nonexistent bit (vacuum-state). This is a 4th bit value, however the vacuum state is the same for all bits thus it is a squeezed logic state [3,4]. BINBL has also a vacuum state that represents a 3rd bit value there however, because it is squeezed, we do not normally count it.

*2.2 Some superpositions in TINBL*

**(e)** Universe without vacuum-states:

$$U_{nv}(t) = \left[ R_{10}(t) + R_{11}(t) + R_{10}(t)R_{11}(t) \right]\left[ R_{20}(t) + R_{21}(t) + R_{20}(t)R_{21}(t) \right] \\ \left[ R_{30}(t) + R_{31}(t) + R_{30}(t)R_{31}(t) \right]...\left[ R_{M0}(t) + R_{M1}(t) + R_{M0}(t)R_{M1}(t) \right] \tag{8}$$

$U_{nv}(t)$ has $3^M$ product strings. Its advantage is that its amplitude is never zero. Thus the disadvantage of other solutions [6,7,11] for avoiding the zero amplitudes of the Universe can be avoided.

**(f)** The total Universe including vacuum-states:

$$U_{tot}(t) = \left[ R_{10}(t) + R_{11}(t) + R_{10}(t)R_{11}(t) + 1 \right]...\left[ R_{M0}(t) + R_{M1}(t) + R_{M0}(t)R_{M1}(t) + 1 \right] \tag{9}$$

$U_{tot}(t)$ has $4^M$ elements in the superposition. Out of all the possible product strings, it contains also the single basic RTW elements, too. Such a universe can be useful in some special-purpose applications (traveling salesmen problems, graph coloring problems, etc.).

*2.3 Examples for logic gates in TINBL*

Here we show a few simple examples to illustrate the differences between the BINBL and TINBL systems. For binary inputs the TINBL gates function in the same way as in the BINBL system.

2.3.1 The NOT/Annihilation/Creation gate

The NOT gate is the same as in BINBL:

$$\text{NOT}_i\left[Y(t)\right] = R_{i0}(t)R_{i1}(t)Y(t) \quad , \tag{10}$$

where $Y(t)$ is the signal of the input state.

The gate works for the binary values of the logic states in the same way, as earlier:

$$\text{NOT}_i\left[H_i\right] \equiv R_{i0}(t)R_{i1}(t)R_{i1}(t) = R_{i0}(t) \equiv L_i \quad , \tag{11}$$

$$\text{NOT}_i\left[L_i\right] \equiv R_{i0}(t)R_{i1}(t)R_{i0}(t) = R_{i1}(t) \equiv H_i \quad . \tag{12}$$

This fact implies that, when the TINBL system is running a BINBL algorithm, the CNOT and XOR gates also function in the same way as earlier because these gates are based on the NOT gate. The same is true for BINBL algorithms with the old gates (e.g. [9,11,13,15,16]).

The NOT gate is an *Annihilation gate* for the relevant bit with the uncertain bit value:

$$\text{NOT}_i\left[X_i\right] \equiv \text{NOT}_i\left[R_{i0}(t)R_{i1}(t)\right] = R_{i0}(t)R_{i1}(t)R_{i0}(t)R_{i1}(t) = 1 \equiv V \quad , \tag{13}$$

where the output value the vacuum $V$ (=1) indicates that the $i$-th bit with uncertain value is removed from the string.

On the other hand, when the bit does not exist in the string, which means its value is a steady 1 amplitude within the product, the NOT gate acts a *Creation gate* as it creates the missing bit and provides an uncertain value for it:

$$\text{NOT}_i[V] \equiv \text{NOT}_i[1] = R_{i0}(t)R_{i1}(t) \equiv X_i \ . \tag{14}$$

2.3.2 The XOR and XNOR (also Creation and Annihilation) gates for single bit output

This versions work only in a single bit system, not on multi-bit product strings:

$$\text{XOR}[A,B] \equiv A(t)B(t)L(t) \ , \tag{15}$$

$$\text{XNOR}[A,B] \equiv A(t)B(t)H(t) \ . \tag{16}$$

The related truth tables are shown in Tables 1 and 2, respectively. It is apparent that these gates can also act as Annihilation and Creation gates when the one or both inputs contain X and/or V values. For classical binary input values they act as the classical XOR and XNOR gates, respectively

|   | **L** | **H** | **X** | **V** |
|---|---|---|---|---|
| **L** | L | H | X | V |
| **H** | H | L | V | X |
| **X** | X | V | L | H |
| **V** | V | X | H | L |

Table 1. Ternary single-output bit XOR gate truth table. Bold: input values. Plain: output values.

|   | **L** | **H** | **X** | **V** |
|---|---|---|---|---|
| **L** | H | L | V | X |
| **H** | L | H | X | V |
| **X** | V | X | H | L |
| **V** | X | V | L | H |

Table 2. Ternary single-output bit XNOR gate truth table. Bold: input values. Plain: output values.

## 3. Conclusion

A three-valued instantaneous noise-based logic is proposed. The third value is an uncertain bit value that can be useful in artificial intelligence applications. There is a forth value, too, that can represent a non-existing bit (vacuum-state) that is the same (1 steady amplitude) for all bits, however that is a squeezed state which is common for all bits. The same vacuum state exists also in binary INBL.

Some simple logic gates are shown to illustrate the differences between the BINBL and TINBL systems. All the known binary logic gates work for the binary bit values in the ternary system in same way as earlier.

The ternary Universe without the vacuum states has an important advantage compared to the standard binary one: its amplitude is never zero during any clock period.

In summary, the former binary algorithms (e.g. 9,11,13,15,16]) can also be run in the ternary system with no change and without the problems posed by zero values of the Universe. Future works will explore the applicability of TINBL to create new logic gates and to solve special purpose problems.

**Acknowledgements**